# Technical Activities Forum

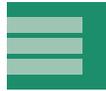

# Cluster Computing: A High-Performance Contender

Mark Baker, University of Portsmouth
Rajkumar Buyya, Monash University
Dan Hyde, Bucknell University

When you first heard people speak of Piles of PCs, the first thing that came to mind may have been a cluttered computer room with processors, monitors, and snarls of cables all around. Collections of computers have undoubtedly become more sophisticated than in the early days of shared drives and modem connections. No matter what you call them—Clusters of Workstations (COW), Networks of Workstations (NOW), Workstation Clusters (WCs), Clusters of PCs (CoPs)—clusters of computers are now filling the processing niche once occupied by more powerful stand-alone machines.

In its simplest form, the computers in your office that are connected to your local area network constitute a workstation cluster. In addition to the hardware, a workstation cluster also includes the middleware that allows the computers to act as a distributed or parallel system and the applications designed to run on it.

While a system based on low-end workstations and network technologies may not at first seem particularly useful, such systems have been the testbeds for a new computing paradigm: high-performance and high-availability cluster computing. This class of system is becoming increasingly commonplace; in fact, most academic institutions and industries that use high-performance computing either already use or are thinking of using workstation clusters to run their most demanding applications. Even companies that can afford traditional supercomputers are becoming interested in commodity clusters.

Why the switch? For some, cluster-based systems provide a way to stretch their computing dollars, allowing the reuse of seemingly obsolete office or classroom systems. Others have found that a cluster of high-performance workstations can easily compete with the best supercomputers IBM or SGI have to offer. A company can download a few tools from a public Web site and order a collection of machines and network equipment to put together an 8-Gflops system for around $50,000. Assembling a powerful supercomputer would cost around $200,000.

## TECHNOLOGIES, COMPONENTS, AND APPLICATIONS

A cluster consists of all the components found on any LAN with PCs or workstations: individual computers with their processors, memory, and disks; network

### Cluster Computing Educational Resources

Although many educational institutions teach undergraduate and graduate students about the hardware and software components that make up a cluster, few courses or programs concentrate on the wealth of technologies that constitute the complete cluster environment, from hardware to application development tools. In order to introduce cluster computing into the curricula of more college programs, the Task Force on Cluster Computing has set up a Web site, http://www.coe.uncc.edu/~abw/parallel/links.html. This informative resource provides links to related journals, books, freely available software, projects from both academia and industry, white papers, and descriptions of hardware components. In addition, with our educational donation program, we actively support academic faculty members around the world who are interested in introducing new cluster-based courses by providing sample curricular materials.

With the generous cooperation of leading publishers worldwide, we have arranged for the donation of some current books on cluster computing. While the books will be available for faculty members who request them, the TFCC has reserved 50 percent for donation to academic programs in developing countries. The titles available include

- *High Performance Cluster Computing: Architectures and Systems*, R. Buyya (ed.), Prentice Hall, 1999
- *High Performance Cluster Computing: Programming and Applications*, R. Buyya (ed.), Prentice Hall, 1999
- *In Search of Clusters*, 2nd ed., G.F. Pfister, Prentice Hall, 1998
- *Metacomputing: Future Generation Computing Systems*, W. Gentzsch (ed.), Elsevier, 1999
- *Parallel Programming: Techniques and Applications Using Networked Workstations and Parallel Computers*, B. Wilkinson and C.M. Allen, Prentice Hall, 1998 cards; cabling; libraries; operating systems; middleware; tools; and various other utilities. The architecture of clusters, however, can vary rather dramatically. At one end of the spectrum are clusters based on commercial off-the-shelf components or put together from older systems, maybe originally used in offices or classrooms. (For information on one of the first COTS clusters, see the Beowulf project home page, http://www.beowulf.org/.) At the other end are proprietary clusters built around high-end SMP processors and custom network technologies. The physical configurations of clusters also vary widely, including anything from a bunch of PCs located in a classroom to motherboards stored in custom racks in a computing services room.

Of course, all these components are there to support applications of one sort or another. Applications appropriate for clusters are not restricted to those traditionally run on supercomputers or other high-performance systems. The number and types of applications now using clusters are increasing all the time. Cluster-based systems support both high- performance parallel applications such as computational chemistry, astrophysics, and computational fluid dynamics, and commercial applications such as a load-balanced high-performance Web server like HotBot, which uses a parallel Oracle database.

## SUPERCLUSTER SYSTEMS AND ISSUES

The National Computational Science Alliance (http://www.ncsa.uiuc.edu/alliance/alliance/), an academic partnership involving more than 50 US universities and research institutions, has built a supercluster consisting of 192 nodes based on dual-processor Pentiums running Windows NT and using Myrinet interconnects. A number of well-known scientific codes have been ported onto the NT supercluster. Its capability has been proven: Several applications attain a sustained performance of around 7 Gflops. This compares favorably to the performance reached on an SGI O2000.

However, as is the case in many professional environments, the differing technologies used in cluster computing can spark debate. For example, many disagree on whether a cluster should use Ethernet—Fast or Gigabit—technology or specialized nonstandard intelligent network cards and protocols like those produced by Myrinet or SCI. Experts in cluster computing circles, like those in other fields, also dispute the best operating system to use: commercial products such as Microsoft's Windows NT or freely available systems such as Linux.

The arguments over operating systems for cluster-based applications, like most arguments, are a combination of objective and subjective reasoning. To settle the OS debate more impartially, research teams working for NCSA have pitted a Linux-based supercluster against one running Windows NT. Housed at the University of New Mexico's Albuquerque High Performance Computing Center, the Linux cluster (http://www.arc.unm.edu/alta/) will run a range of computationally intensive applications to be compared to the proven power of the NCSA's NT cluster (http://www.ncsa.uiuc.edu/General/CC/ntcluster/).

## THE NEED FOR A NEW TASK FORCE

Recognizing the trend toward clusters for high-performance computing, the IEEE Computer Society has approved a Task Force on Cluster Computing (TFCC). You may ask, "What's so special about that? People have been using clusters of computers for years. It's a bit like Microsoft realizing that the Internet may be a big thing." But think again. The overwhelming number of cluster-related projects and products appearing in the development arena and the commercial marketplace means that a focused group can help lead international efforts in cluster-based computing.

Why not place cluster-based activities under the umbrella of another related Technical Committee, such as the committees on the Internet, Supercomputing Applications, or Distributed or Parallel Processing? After all, what is cluster computing but a mixture of these disciplines? But cluster computing combines so many computing concepts and technologies that placing it under an existing banner would dilute the attentions of individuals interested in all the aspects that come together in this field.

With the advent of the TFCC, interested Computer Society members can participate in one focused group to champion the cause of cluster computing by sponsoring workshops, conferences, projects, and standards. The TFCC has set up two mirrored Web sites: one in Australia (http://www.dgs.momash.edu.au/~rajkumar/tfcc/), and another in the UK (http://www.dcs.port.ac.uk/~mab/tfcc/), as media for timely communication of its activities.

---

### Cluster Computing Workshop Scheduled for December

Clusters of computers are rapidly catching up with the processing capabilities of more costly supercomputers, traditional parallel processors, and proprietary mission-critical systems. To share the latest findings on building efficient clusters, the Task Force on Cluster Computing is planning the First IEEE Computer Society International Workshop on Cluster Computing.

Scheduled for 2 December in Melbourne, Australia, IWCC will take place following PART 99, the sixth Autralasian Conference on Parallel and Real-Time Systems. Spanning the issue of cluster computing from hardware to middleware and applications, IWCC will include presentations on communication protocols, tools for operating and managing clusters, job and resource management, data distribution, load balancing, and programming environments for clusters.

See http://www.dgs.monash.edu.au/~rajkumar/tfcc/IWCC99/ for more details on IWCC. Proceedings from the workshop will be available from the Computer Society at http://computer.org/conferen/proceed/dlproceed.htm. Tentative registration fees for IWCC are $175 for members, $200 for nonmembers, and $150 for full-time students. The registration fee for students is likely to be reduced: We are expecting the donation of sponsorship funds from businesses.

### Thompson Receives Kanai Award in June Ceremony

In a Richmond, Virginia, awards ceremony in June, Kenneth L. Thompson of Lucent Technologies was presented with the first IEEE Computer Society Tsutomu Kanai Award for his role in creating Unix. Society President Leonard L. Tripp presented Thompson with the crystal memento and $10,000 honorarium, made possible through an endowment from Hitachi Ltd., where Tsutomu Kanai served as president for 30 years.

To find Thompson's views on the state of computing today, read "Unix and Beyond: An Interview with Ken Thompson" by Daniel Cooke, Joseph Urban, and Scott Hamilton, *Computer*, May 1999 pp. 58-64.

### What's New in Computer Society Periodicals

**SOFTWARE** Want to know what's happening in software certification? Read the July-August issue of *IEEE Software.* The issue also features a debate on the future of Linux.

**Computer Graphics** Learn about the use of color in computer graphics, especially to enhance communication, in a general-interest tutorial in the July-August issue of *IEEE Computer Graphics and Applications.*

**Internet Computing** The July-August issue of *IEEE Internet Computing* examines emerging protocols such as dynamic host configuration, service location, and ad hoc networking for simplifying network autoconfiguration.

**Design&Test** For in-depth coverage on testing and the product life cycle, read the July-September issue of *IEEE Design & Test of Computers,* published to coincide with the 1999 International Test Conference, 26-30 September, in Atlantic City, N.J.

**Intelligent SYSTEMS** The May-June issue of *IEEE Intelligent Systems,* which features self-adaptive software, spotlights a computer program that can write music in a particular composer's style. The issue also marks the debut of a new department, *AI in Space.*

**MICRO** *IEEE Micro's* May-June issue features articles on identifying design bugs. Also in this issue: an inside look at the development of Sun Microsystems' UltraSPARC-III.

**MultiMedia** *IEEE MultiMedia* devotes its April-June issue to media spaces—the convergence of computing, communication, and media, with applications in teleconferencing and other areas.

Computer Society periodicals and proceedings are available online to members who have access to the digital library (http://computer.org/epub/), available until 15 August for $50. Individual copies of back issues are available from cs.books@computer.org.

---

**Technical Activities Forum**

### INTERNATIONAL ACTIVITIES

The TFCC, like other Society activities, is supported by a group of enthusiastic international volunteers from both academia and industry. More than 90 scholars from around the world are involved in the management of the Task Force.

The TFCC plans to hold a number of workshops, including invited talks and submitted paper and panel sessions that span a wide range of cluster-based technologies. Two events are already being planned. The International Workshop on Cluster Computing (see the sidebar on p. 80, "Cluster Computing Workshop Scheduled for December"), cosponsored by the TFCC, the IEEE Computer Society, the Asian Technology Information Program, MPI Software Technology Inc., and GENIAS Software Inc., will be held on 2 December in Melbourne, Australia. The TFCC will also have a presence at SC99, 13-19 November, in Portland, Oregon. Further events for 2000 are in the pipeline, and we plan to start an annual international cluster computing series.

The TFCC will also hold tutorials, workshops, and symposia in conjunction with existing conferences hosted by the IEEE and the Computer Society to help increase participation and cooperation among other relevant Society Technical Committees. In addition, the TFCC publishes a biannual newsletter to keep its members abreast of the events, initiatives, and latest developments within this field.

### Industrial promotion

Cluster-based systems are widely used in the commercial sector. Recognizing the importance of close cooperation and coordination between academia and industry, the TFCC established an industry focus group to strengthen research and development ties between the two sectors. This group also seeks to find potential collaboration opportunities by monitoring the activities of other international research groups and standardization bodies.

---

The Task Force on Cluster Computing will only flourish with the aid of active members. You can read about our efforts by visiting our Web site or joining our open TFCC mailing group. Details on how to subscribe to the mailing list are given on our Web site. We encourage Society members to be a part of TFCC activities. To sign up, fill out the form at http://computer.org/tcsignup/. ❖

*Mark Baker* and *Rajkumar Buyya* *are co-chairs of the Task Force on Cluster Computing. Contact them at Mark.Baker@computer.org and rajkumar@dgs.monash.edu.au. Dan Hyde is the TFCC newsletter editor. Contact him at hyde@bucknell.edu.*